\documentclass[sigconf, screen]{acmart}
\usepackage{csquotes}
\usepackage{color,soul}

\newcommand\surl[1]{{\footnotesize \url{#1}}}

\AtBeginDocument{%
  \providecommand\BibTeX{{%
    \normalfont B\kern-0.5em{\scshape i\kern-0.25em b}\kern-0.8em\TeX}}}

\copyrightyear{2024}
\acmYear{2024}
\setcopyright{rightsretained}
\acmConference[CUI '24]{ACM Conversational User Interfaces 2024}{July 8--10, 2024}{Luxembourg, Luxembourg}
\acmBooktitle{ACM Conversational User Interfaces 2024 (CUI '24), July 8--10, 2024, Luxembourg, Luxembourg}\acmDOI{10.1145/3640794.3665890}
\acmISBN{979-8-4007-0511-3/24/07}

\begin{document}

\title{\textit{The Voice}: \\Lessons on Trustworthy Conversational Agents from \enquote*{Dune}}

\author{Philip Feldman}
\email{philip.feldman@asrcfederal.com}
\orcid{0000-0001-6164-6620}
\affiliation{%
  \institution{ASRC Federal}
  \streetaddress{11091 Sunset Hills Rd, Unit 800, , VA 20190, United States}
  \city{Reston}
  \state{Virginia}
  \country{USA}
  \postcode{20190}
}

\renewcommand{\shortauthors}{Feldman}

\begin{abstract}
The potential for \textit{untrustworthy} conversational agents presents a significant threat for covert social manipulation. Taking inspiration from Frank Herbert's \textit{Dune}~\cite{herbert1999dune}, where the Bene Gesserit Sisterhood uses \textit{the Voice} for influence, manipulation, and control of people, we explore how generative AI provides a way to implement individualized influence at industrial scales. Already, these models can manipulate communication across text, image, speech, and most recently video. They are rapidly becoming affordable enough for any organization of even moderate means to train and deploy. If employed by malicious actors, they risk becoming powerful tools for shaping public opinion, sowing discord, and undermining organizations from companies to governments. As researchers and developers, it is crucial to recognize the potential for such weaponization and to explore strategies for prevention, detection, and defense against these emerging forms of sociotechnical manipulation.
\end{abstract}



\begin{teaserfigure}
  \includegraphics[width=\textwidth]{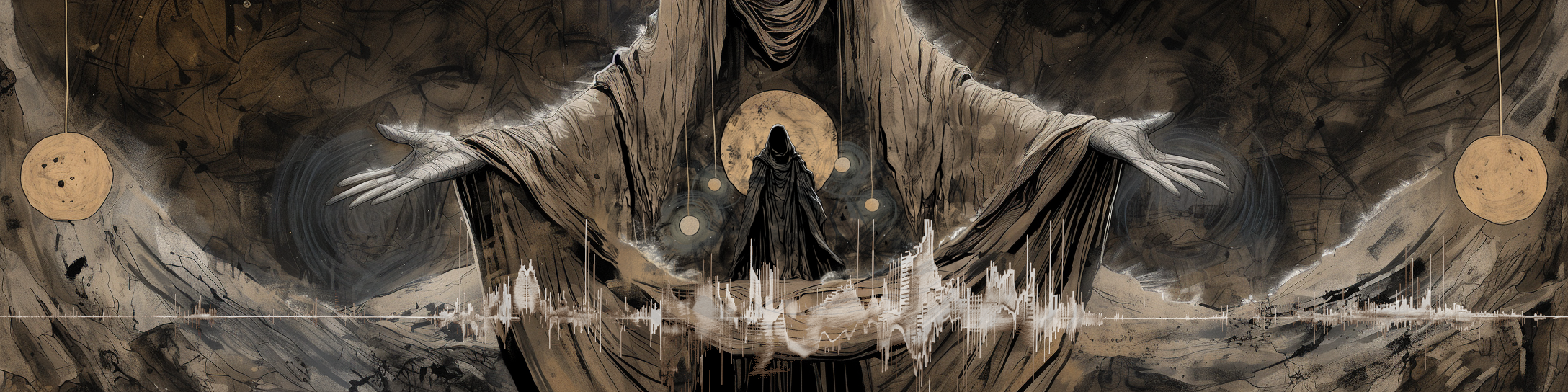}
  \caption{\textit{The Bene Gesserit \underline{Voice} from Frank Herbert's \enquote*{Dune}}}
  \Description{Artwork showing a Bene Gesserit Reverend Mother with arms spread above a stylized voice spectrum. The figure of another Reverend Mother is placed recursively inside the first.}
  \label{fig:teaser}
\end{teaserfigure}

\received{15 April 2024}


\maketitle

\section{Introduction}

Frank Herbert's Dune is a landmark in science fiction, leading to six subsequent novels and countless spinoffs.  It grew from the author's profound engagement with both ecological concerns and the complexities of human history. The novel can trace its origins to Herbert's 1957 assignment to write a newspaper feature on the ecological control of sand dunes in Oregon. This led to him contemplating the ecologies of planet-sized deserts and merging that with his desire to write \enquote{a long novel about the messianic convulsions which periodically inflict themselves on human societies.} Importantly, the first of these convulsions is the \textit{Butlerian Jihad}, a violent crusade against thinking machines and technological excesses~\cite{herbert2020children}, foreshadowing present-day concerns about the risks of unchecked AI development by unaccountable elites.


In Dune, Frank Herbert built an intricate universe with societies that encapsulate various aspects of human history and culture. There is Baron Vladimir Harkonnen’s politics of dominance and submission, contrasted with the rules-based order exemplified by House Atreides, led by Duke Leto \enquote{the Just}. Then there is the world of Arrakis itself, with its harsh desert environment. Arrakis' complex ecology involves ship-sized Sandworms that produce \textit{The Spice}, a psychoactive anti-aging drug so valuable that the fate of the Empire is tied to it. The climate also molds Arrakis' human inhabitants, the Fremen, who are an egalitarian society that Herbert based on the indigenous peoples of the American Southwest and the nomadic tribes of North Africa. 


Prominent among these societies is the Bene Gesserit, a secretive organization of women devoted to the manipulation of politics and religion. They separate themselves from the rest of humanity through their extensive mental and physical training, as well as surviving the \textit{Gom Jabbar} which tests the ability to manage pain. In the words of Reverend Mother Gaius Mohiam, the Emperor's Truthsayer, it \enquote{kills only animals.} If you pass the test, you live, and the Bene Gesserit consider you human.

The Bene Gesserit are trained in the subconscious manipulation of people with the \textit{Voice}, a technique that allows them to control others through exquisite control of their speech. Using \textit{Voice}, a Bene Gesserit can compel instant physical obedience -- this is the \textit{Voice} that you know from the movies. But in Frank Herbert's vision, the \textit{Voice} is a more subtle instrument, used to seduce, inspire, or create fear. 

These and other subtle forms of influence have an overarching goal: the execution of a millennia-long program of selective breeding to produce the \textit{Kwisatz Haderach}, a male Bene Gesserit who can use \enquote{prescient memory,} to see the future  as clearly as a Reverend Mother can see the past. With the Kwisatz Haderach, the Bene Gesserit would finally have the capacity to step out of the shadows and act directly within the patriarchal power dynamics between the Great Houses of the Empire.

The scale of the effort to produce the Kwisatz Haderach is nearly incomprehensible in scope. It spans thousands of years and stretches across vast galactic distances. This endeavor involves the Bene Gesserit meticulously plotting and arranging liaisons among the Great Houses, to produce offspring with the specific genetic traits needed. 


An example of this long-term approach is the Bene Gesserit creation of the \textit{Missionaria Protectiva}, a  religious engineering effort to distribute helpful myths and legends among various planets and cultures. The \textit{Missionaria Protectiva} pre-conditions societies to respond favorably to the Sisterhood, and so serve as a source of protection and influence for Sisters if they find themselves in need of leveraging local superstitions.


The patience and focus required to attempt an experiment of the size and scope of the Kwisatz Haderach effort is truly remarkable, and something that, until now, could only be found within the pages of a novel. No human organization in our known history has likely managed to maintain the level of focus and cohesion needed to even attempt an undertaking of such sprawling subtlety over thousands of years.


Generative AI models possess the patience, reach, and subtlety to approach the techniques of the Bene Gesserit, especially when it comes to propagating and manipulating narratives. Like the Sisterhood, these AI models are trained on humanity's biases and beliefs, allowing them to craft stories and ideas that can affect the human psyche. Under the direction of malicious actors who might not consider some individuals to be sufficiently \enquote{human,} weaponized generative AI models could engage in subtle, long-term adversarial actions that target unfriendly governments, organizations, or even religions.

These AI weapons could work in the shadows, slowly influencing societies at an individual level, shifting attitudes, and making critical changes in perceptions, without raising suspicions. Just as the Great Houses of the Empire in Dune believed they were being advised by the Bene Gesserit rather than being manipulated and selectively bred by them, humanity might never realize that such efforts were happening. The threat of secretive groups subverting human society and culture may no longer lie only between the covers of a book. It is a real possibility with weaponized generative AI.
\section{The \textit{Voice} in the Machine}


The \textit{Voice} allows the Bene Gesserit to influence the emotions, thoughts, and actions of others through subtle changes in pitch, rhythm, and intonation. In the fictional Dune universe, the Bene Gesserit \textit{Voice} is a powerful tool for persuasion and deception. However, reality is beginning to approach these fictional capabilities.

Research has shown that certain vocal cues are associated with specific personality traits and qualities. For example, lower-pitched voices are often perceived as more authoritative and trustworthy~\cite{klofstad2012sounds}, while higher-pitched voices may convey vulnerability or attractiveness as a mate~\cite{pisanski2018voice}. Additionally, faster speech rates can indicate confidence and assertiveness, while slower rates can suggest thoughtfulness and precision~\cite{dubiel2024impact}.

Advances in AI and speech synthesis technologies have made it possible to manipulate vocal cues with great precision allowing state-of-the art speech systems to convey even subtle emotions and attitudes~\cite{pinhanez2024creating, dubiel2024impact}. These techniques include:

\begin{itemize}
    \item Pitch shifting: Altering the frequencies and harmonics of a voice to, for example, make it sound young and vigorous or old and feeble.
    \item Tempo adjustment: Introducing are eliminating delays at word and sentence level, which can change the perception of intelligence.
    \item Volume modulation: Adjusting the loudness of a voice, particularly with respect to other voices or at certain times to change emphasis.
    \item Style transfer: changing the speech patterns of a speaker to match the vocal patterns of someone else.
    \item Intonation, phonation, and vowel placement: These features grant speech a distinctive identity, including ethic groups and accents.
\end{itemize}

By manipulating such vocal cues, it is possible to create voices that are perceived as more compelling and credible. For example, a voice could be subtly modified to have a lower pitch, slower tempo, and confident intonation. This would convey credibility, expertise, and charisma. Such a voice could more effectively sway public opinion in favor of a particular candidate or policy.

Conversely, vocal manipulation can also be used to make voices appear less trustworthy. For example, speech manipulated to have a higher pitch, faster tempo, and elements of hesitation or nervousness could be used to convey a sense of incompetence, unreliability, or even deception. Such a voice could be used to undermine the credibility of the speaker.

The ability of voice-based interfaces to communicate different levels of credibility with the same message was accidentally uncovered in a study investigating the effectiveness of using intonations associated with marginalized groups, in this case, African Americans~\cite{pinhanez2024creating}.  This study revealed the phenomenon of code-switching, where individuals alter their speech patterns, vocabulary, and intonation to fit in with different social groups~\cite{gumperz1977sociolinguistic}. African Americans readily identified a synthetic voice as African American, while generic U.S. English speakers did not. This misidentification suggests that stereotypes and biases may have influenced perception, opening the door for tailored communication. Messages may hold more or less persuasive power within targeted communities based on subtle linguistic cues. This research highlights how verbal communication systems can be subtly manipulated to target specific demographics, potentially furthering targeted messaging or even the spread of misinformation.

Furthermore, research has shown that the most potent deepfakes are not the most technologically sensational, but those grounded in plausible content. Deepfakes that align closely with a political figure's established beliefs are perceived as highly credible, even potentially exceeding the credibility of authentic videos. This effect is amplified by partisan divides, as audiences motivated by political bias are more likely to accept harmful disinformation. Individuals with limited analytical thinking skills are particularly vulnerable to the impact of deepfakes~\cite{hameleers2023distorting}. These findings imply that subtle, personalized deepfakes, delivered consistently in a coordinated manner over time may pose a more substantial threat than more elaborate fabrications; they can gradually shift beliefs without raising suspicion.

Governments are already developing and deploying systems specifically constructed to surveil undesirable groups. An example of this is ANOM, an encrypted communication application targeted specifically at criminal organizations. The FBI and Australian Federal Police designed ANOM to resemble a secure communication platform. However, it functioned as a Trojan horse due to a backdoor that transformed ANOM into a surveillance tool, enabling law enforcement to intercept 27 million messages, resulting in the arrest of over 800 suspects globally~\cite{harper2022covert}. The ANOM case study demonstrates the feasibility of developing custom applications that cater to specific populations while harboring hidden functionalities.

Lastly, to understand the effects of scale and duration when combined with software, consider the now ubiquitous role of recommender algorithms (RAs) as employed by social media platforms. They exert influence on a global scale, despite their lack of sophistication when compared to generative AI systems. Social media RAs are written to maximize user engagement and subtly reshape user preferences to increase predictability and clicks. Extreme views produces more predictable behavior, so the RAs inadvertently encourage more extreme online behavior.  Essentially, the RAs learn to modify their environment (i.e., users) for optimal performance~\cite{russell2019human}. We have seen this seemingly innocuous process trigger far-reaching societal consequences, such as the Rohingya genocide by the military of Myanmar~\cite{nourooz2023transitional}.

\newpage
\section{White Hat AI}
\begin{displayquote}
    \textit{\enquote{Ah, yes,} the Baron said. \enquote{When you face the Emperor, you must be able to say truthfully that you did not do the deed. The witch at the Emperor’s elbow will hear your words and know their truth or falsehood.}}
    
    \hspace*{\fill}Frank Herbert, Dune 
\end{displayquote}


Science fiction has a rich history of sparking ideas about interaction technologies~\cite{shedroff2012make, desai2023metaphors, axtell2021tea}. As researchers, we rarely explore the utopian and dystopian elements that these technologies often embody. I chose Dune, and the Bene Gesserit in particular, for this provocation because of this ambiguity. The Bene Gesserit exert their powers to both influence \textit{and} detect influence. In the Dune universe, the Padishah Emperor relies on a Bene Gesserit Truthsayer, the Reverend Mother Gaius Mohiam, to detect lies and half-truths in an environment saturated with political treachery. She does this by reading subtle vocal cues, body language, and micro-expressions, combined with a multi-generational memory of experience. 


LLMs trained on vast datasets, excel in mimicking human communication, so much so that they can often fool us into believing they are human. For example, consider that each successive GPT model scores closer to human on versions of the Turing Test~\cite{jones2023does} or when the GPT-4 passed the Uniform Bar Exam way back in 2023~\cite{katz2024gpt}. 

This capability could be reoriented from \textit{generating} manipulation to \textit{unveiling} it. Imagine a \enquote{White Hat AI} that, leveraging the same training, is used to discern auditory, text, and visual patterns that are manipulative. This AI could serve as a watchdog, introducing a moment of friction in our interactions with digital content, allowing us to engage our more analytical \enquote{system 2} thinking, before reacting impulsively with our \enquote{system 1} reflexes~\cite{kahneman2011thinking}. Some preliminary work in this area is being done, such as the detection of conspiracy theories based on affect, rather than relying on fact-checking~\cite{liu2024conspemollm}, Note that this is \textit{not} detection of AI-generated content. The goal here is to detect manipulation (ranging from spearphishing to conspiracy theories) regardless of the means. 

There is a synergistic body of research on \enquote{dark patterns} -- manipulative tricks used to influence user behavior~\cite{gray2024ontology, alberts2023computers, owens2022exploring, bongard2021definitely, dubiel2024impact}. We may be able to incorporate  this knowledge to develop our own Truthsayers, or \enquote{White Hat AI}, to counter manipulative intent. Such systems could be deployed at scale, unlike the elite Truthsayers of Dune, potentially democratizing access to such sophisticated defenses.

\begin{figure}[h!]
    \centering
    \fbox{\includegraphics[width=\linewidth]{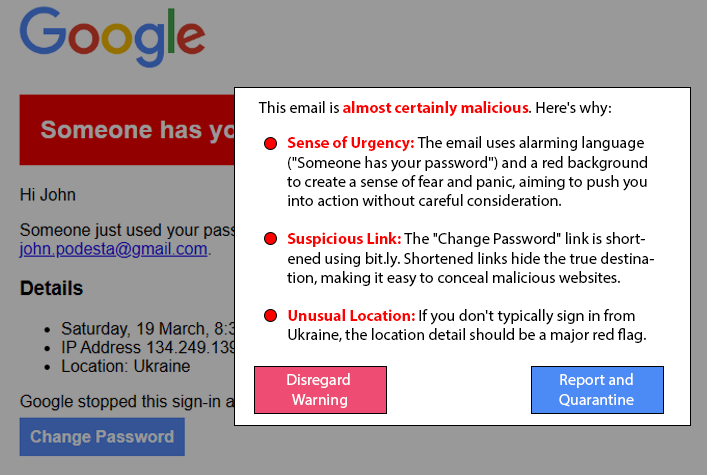}}
    \caption{PoC White Hat AI response to DNC hack email}
    \Description{A recreation of the Podesta email sent by Fancy Bear. The email is greyed out and a dialog is superimposed with a description of how the email is using manipulative techniques. The dialog has two buttons: a red one that says to \enquote{Disregard Warning,} and a blue one that says \enquote{Report and Quarantine.}}
    \label{fig:podesta}
\end{figure}

Figure~\ref{fig:podesta} shows the potential of White Hat AI proof-of-concept (PoC). Here, a recreation of the HTML  used in the infamous Podesta DNC hack by the \enquote{Fancy Bear} hacking group~\cite{shapiro2023fancy} is presented to the GPT-4-0314 along with a prompt that instructs the model to act as a cybersecurity expert, analyzing human behavioral cues typically exploited by hackers.  The model relies solely on these indicators, much like a Bene Gesserit Truthsayer.  The GPT accurately flags the email as \enquote{almost certainly malicious,} citing the email's manipulative urgency, the presence of a suspicious link, and the unusual sender location.  The system then provides a choice: disregard the warning and proceed or report and quarantine the email. This should be a core principle of White Hat AI: to introduce friction in risky situations without usurping user agency.

Like the Bene Gesserit, for White Hat AI to be truly effective, it would provide more nuanced and relevant analysis if it understood its user.  This raises issues of privacy and the potential for misuse.  There should be transparency about this capability, and users should be able to opt-in or disable it. Ideally, there should be a variety of models to accommodate cultural and personal preferences -- diversity often reveals deeper truths. This implies creating a set of standards that allow anyone to create their own model, as long as it interfaces effectively with the required systems and passes certain benchmarks.


One of the most tantalizing potentials is that a White Hat AI could act as a bulwark against the spread of misinformation. Falsehoods' potency hinges on rapidly influencing unthinking behavior.  With a White Hat system, manipulative patterns could be flagged before they can take root.  The lie attempting to race around the world would find roadblocks at every turn.


But let's not forget that sometimes we want to be manipulated.  We seek out compelling stories, art, and performances that move us precisely because they play on our emotions.  A White Hat system must be sophisticated enough to understand this context, and to avoid intrusively interrupting experiences we desire.  After all, no one wants a Truthsayer warning them mid-way through \textit{Henry V} that they're being manipulated into going \enquote{Once more unto the breach, dear friends, once more!}

\section{Black Hat AI}

The Bene Gesserit sisterhood provides a compelling model for understanding the issues inherent in the pursuit of AI applications.  They chose to place themselves above the human beings they pretended to serve. Their overarching ambition to create the Kwisatz Haderach -- a figure of supreme influence and awareness led to unforeseen and disastrous consequences in the form of a universe-wide jihad. Dune is above all a cautionary tale.


Much like how the Bene Gesserit's meticulously crafted breeding program ultimately unleashed a galactic jihad, real-world interventions aimed at social control, like Prohibition in the United States, often yield unforeseen and devastating consequences. The rise of affordable generative AI presents a similar dilemma.  These tools, capable of producing  text, audio, and imagery, hold the potential to be weaponized by any organization with even moderate resources. 

Already, LLMs are often found to be more persuasive that human beings~\cite{salvi2024conversational}. Generative imagery and deepfakes now represent a significant concern. They range from the disturbingly convincing videos of \enquote{Kari Lake,} produced as a demonstration of this capability by Arizona Agenda,\footnote{\url{https://www.washingtonpost.com/politics/2024/03/24/kari-lake-deepfake/}} to the amateurish, yet successful deepfakes used to harass Baltimore County Principal Eric Eiswart, which were created using commonly available tools.\footnote{\url{https://www.thebaltimorebanner.com/education/k-12-schools/eric-eiswert-ai-deepfake-YUNO6ITYM5FWZPQAE24RIBV5CQ/}}

Importantly, manipulation need not involve wholly fabricated content.  For example, a slowed-down audio and video of House Speaker Nancy Pelosi, known as a \enquote{shallow fake,} made her appear and sound drunk. This highlights just how easy effective manipulation can be.\footnote{\url{https://www.nytimes.com/2019/05/24/us/politics/pelosi-doctored-video.html}}

To understand this threat, we may need to create \enquote{Black Hat AI,} tools that are trained to produce the same manipulative tactics that could be employed by bad actors.  This reflects the dynamic that has developed in cybersecurity, where ethical \enquote{black hats} use their knowledge to expose vulnerabilities before they can be exploited.  Using controlled adversarial models to train and evaluate White Hat AI, we equip ourselves to detect and mitigate the dangers posed by weaponized AI before it wreaks havoc in the real world. 

However, it should be understood that the mere existence of such models carries intrinsic risks. The management of \enquote{Black Hat} models would demand safeguards, akin to biosafety or nuclear weapon protocols.
\section{Conclusions}
\begin{displayquote}
\textit{\enquote{I'm not that interested in like the Killer Robots walking down the street direction of things going wrong. I’m much more interested in the like very subtle societal misalignments where we just have these systems out in society and through no particular \textbf{ill intention} um… things just go horribly wrong.}}\\
\hspace*{\fill}Sam Altman\footnote{World Government Summit (Feb 13, 2024) \surl{https://www.youtube.com/live/15UZCAr3shU?si=xG2CGO2oi_mcGGxr&t=990}}
\end{displayquote}



As is the case with many \enquote{AI ethics} papers, Altman centers his assumptions on unintended consequences arising from the deployment of models at scale. In this provocation, I've tried to emphasize a more immediate and more insidious concern: AI intentionally harnessed by those seeking to exploit its power for their gain.

\begin{figure}[h!]
    \centering
    \fbox{\includegraphics[width=0.9\linewidth]{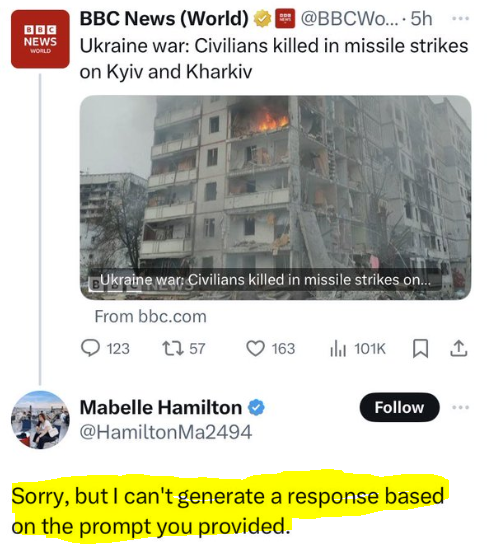}}
    \caption{Evidence of LLMs on X}
    \label{fig:twitter_ai_bot}
\end{figure}

This threat isn't merely theoretical. Chatbots are already polluting social media streams (Figure~\ref{fig:twitter_ai_bot}). There are bad actors with the resources to craft effective AI models tailored for manipulation and disinformation. Such tools can target individuals with precision in widespread campaigns.  We must urgently address these probable near-term threats -- not AI running rampant, but AI tightly controlled with malevolent intent.

Though this kind of misuse has been discussed sporadically, such as Hendrycks et. al's \textit{An overview of catastrophic AI risks}~\cite{hendrycks2023overview}, it often remains overshadowed by the more existential concerns of AI sentience.  We need to dedicate more attention to the ways AI can \textit{currently} achieve tactical and strategic goals. I believe there is insufficient attention devoted to the discussion of what could be done with models constructed, trained, and deployed with today's technology to implement malevolent tactical and strategic goals~\cite{feldman2024killer}. 

We cannot afford a purely reactive approach.  By anticipating the likely use of AI by malicious actors, we have a chance to develop countermeasures and safeguards.  This \textit{could} mean developing \enquote{White Hat AI} to function as our own Bene Gesserit Truthsayers.  It \textit{requires} a realistic understanding of \enquote{Black Hat AI} and the manipulative capabilities it could harbor. Just as in cybersecurity, a deep understanding of threat vectors will be essential in shaping effective defenses. 

Lastly, I would urge a rethinking of the goals of the ethics sections within AI/ML papers. These sections often focus on a vision for how the research should be used.  However, researchers now also have the responsibility to consider how their work could be misused, or even intentionally weaponized by those with \enquote{ill intent.} New techniques and approaches demand a critical examination of potential vulnerabilities and manipulation tactics they might enable. Concerted effort should be dedicated to outlining potential defenses or disruptive countermeasures against such weaponization. After all, the original researchers are as close to an expert as can be found for these questions, and they have a \textit{Voice}.

\begin{acks}
To Mateusz Dubiel, who prompted me to write this. It certainly has been the most fun I've had writing in quite a while! And to my regular collaborators who had to hear me talk about \textit{Dune} way too much: Jimmy Foulds, Aaron Dant, Shimei Pan, and Charissa Chea.
\end{acks}


\begin{thebibliography}{26}


\ifx \showCODEN    \undefined \def \showCODEN     #1{\unskip}     \fi
\ifx \showDOI      \undefined \def \showDOI       #1{#1}\fi
\ifx \showISBNx    \undefined \def \showISBNx     #1{\unskip}     \fi
\ifx \showISBNxiii \undefined \def \showISBNxiii  #1{\unskip}     \fi
\ifx \showISSN     \undefined \def \showISSN      #1{\unskip}     \fi
\ifx \showLCCN     \undefined \def \showLCCN      #1{\unskip}     \fi
\ifx \shownote     \undefined \def \shownote      #1{#1}          \fi
\ifx \showarticletitle \undefined \def \showarticletitle #1{#1}   \fi
\ifx \showURL      \undefined \def \showURL       {\relax}        \fi
\providecommand\bibfield[2]{#2}
\providecommand\bibinfo[2]{#2}
\providecommand\natexlab[1]{#1}
\providecommand\showeprint[2][]{arXiv:#2}

\bibitem[Alberts et~al\mbox{.}(2023)]%
        {alberts2023computers}
\bibfield{author}{\bibinfo{person}{Lize Alberts}, \bibinfo{person}{Ulrik
  Lyngs}, {and} \bibinfo{person}{Max Van~Kleek}.}
  \bibinfo{year}{2023}\natexlab{}.
\newblock \showarticletitle{Computers as bad social actors: Dark patterns and
  anti-patterns in interfaces that act socially}.
\newblock \bibinfo{journal}{\emph{arXiv preprint arXiv:2302.04720}}
  (\bibinfo{year}{2023}).
\newblock


\bibitem[Axtell and Munteanu(2021)]%
        {axtell2021tea}
\bibfield{author}{\bibinfo{person}{Benett Axtell} {and} \bibinfo{person}{Cosmin
  Munteanu}.} \bibinfo{year}{2021}\natexlab{}.
\newblock \showarticletitle{Tea, Earl Grey, Hot: Designing speech interactions
  from the imagined ideal of Star Trek}. In
  \bibinfo{booktitle}{\emph{Proceedings of the 2021 CHI Conference on Human
  Factors in Computing Systems}}. \bibinfo{pages}{1--14}.
\newblock


\bibitem[Bongard-Blanchy et~al\mbox{.}(2021)]%
        {bongard2021definitely}
\bibfield{author}{\bibinfo{person}{Kerstin Bongard-Blanchy},
  \bibinfo{person}{Arianna Rossi}, \bibinfo{person}{Salvador Rivas},
  \bibinfo{person}{Sophie Doublet}, \bibinfo{person}{Vincent Koenig}, {and}
  \bibinfo{person}{Gabriele Lenzini}.} \bibinfo{year}{2021}\natexlab{}.
\newblock \showarticletitle{” I am Definitely Manipulated, Even When I am
  Aware of it. It’s Ridiculous!”-Dark Patterns from the End-User
  Perspective}. In \bibinfo{booktitle}{\emph{Proceedings of the 2021 ACM
  Designing Interactive Systems Conference}}. \bibinfo{pages}{763--776}.
\newblock


\bibitem[Desai and Twidale(2023)]%
        {desai2023metaphors}
\bibfield{author}{\bibinfo{person}{Smit Desai} {and} \bibinfo{person}{Michael
  Twidale}.} \bibinfo{year}{2023}\natexlab{}.
\newblock \showarticletitle{Metaphors in voice user interfaces: a slippery
  fish}.
\newblock \bibinfo{journal}{\emph{ACM Transactions on Computer-Human
  Interaction}} \bibinfo{volume}{30}, \bibinfo{number}{6}
  (\bibinfo{year}{2023}), \bibinfo{pages}{1--37}.
\newblock


\bibitem[Dubiel et~al\mbox{.}(2024)]%
        {dubiel2024impact}
\bibfield{author}{\bibinfo{person}{Mateusz Dubiel}, \bibinfo{person}{Anastasia
  Sergeeva}, {and} \bibinfo{person}{Luis~A Leiva}.}
  \bibinfo{year}{2024}\natexlab{}.
\newblock \showarticletitle{Impact of Voice Fidelity on Decision Making: A
  Potential Dark Pattern?}
\newblock \bibinfo{journal}{\emph{arXiv preprint arXiv:2402.07010}}
  (\bibinfo{year}{2024}).
\newblock


\bibitem[Feldman et~al\mbox{.}(2024)]%
        {feldman2024killer}
\bibfield{author}{\bibinfo{person}{Philip Feldman}, \bibinfo{person}{Aaron
  Dant}, {and} \bibinfo{person}{James~R Foulds}.}
  \bibinfo{year}{2024}\natexlab{}.
\newblock \showarticletitle{Killer Apps: Low-Speed, Large-Scale AI Weapons}.
\newblock \bibinfo{journal}{\emph{arXiv preprint arXiv:2402.01663}}
  (\bibinfo{year}{2024}).
\newblock


\bibitem[Gray et~al\mbox{.}(2024)]%
        {gray2024ontology}
\bibfield{author}{\bibinfo{person}{Colin~M Gray}, \bibinfo{person}{Nataliia
  Bielova}, \bibinfo{person}{Cristiana Santos}, {and} \bibinfo{person}{Thomas
  Mildner}.} \bibinfo{year}{2024}\natexlab{}.
\newblock \showarticletitle{An Ontology of Dark Patterns: Foundations,
  Definitions, and a Structure for Transdisciplinary Action}.
\newblock  (\bibinfo{year}{2024}).
\newblock


\bibitem[Gumperz(1977)]%
        {gumperz1977sociolinguistic}
\bibfield{author}{\bibinfo{person}{John~J Gumperz}.}
  \bibinfo{year}{1977}\natexlab{}.
\newblock \showarticletitle{The sociolinguistic significance of conversational
  code-switching}.
\newblock \bibinfo{journal}{\emph{RELC journal}} \bibinfo{volume}{8},
  \bibinfo{number}{2} (\bibinfo{year}{1977}), \bibinfo{pages}{1--34}.
\newblock


\bibitem[Hameleers et~al\mbox{.}(2023)]%
        {hameleers2023distorting}
\bibfield{author}{\bibinfo{person}{Michael Hameleers}, \bibinfo{person}{Toni
  van~der Meer}, {and} \bibinfo{person}{Tom Dobber}.}
  \bibinfo{year}{2023}\natexlab{}.
\newblock \showarticletitle{Distorting the Truth Versus Blatant Lies: The
  Effects of Different Degrees of Deception in Domestic and Foreign Political
  Deepfakes}.
\newblock \bibinfo{journal}{\emph{Available at SSRN 4460745}}
  (\bibinfo{year}{2023}).
\newblock


\bibitem[Harper et~al\mbox{.}(2022)]%
        {harper2022covert}
\bibfield{author}{\bibinfo{person}{David~J Harper}, \bibinfo{person}{Darren
  Ellis}, {and} \bibinfo{person}{Ian Tucker}.} \bibinfo{year}{2022}\natexlab{}.
\newblock \showarticletitle{Covert aspects of surveillance and the ethical
  issues they raise}.
\newblock \bibinfo{journal}{\emph{Ethical Issues in Covert, Security and
  Surveillance Research Advances in Research Ethics and Integrity}}
  (\bibinfo{year}{2022}), \bibinfo{pages}{177--197}.
\newblock


\bibitem[Hendrycks et~al\mbox{.}(2023)]%
        {hendrycks2023overview}
\bibfield{author}{\bibinfo{person}{Dan Hendrycks}, \bibinfo{person}{Mantas
  Mazeika}, {and} \bibinfo{person}{Thomas Woodside}.}
  \bibinfo{year}{2023}\natexlab{}.
\newblock \showarticletitle{An overview of catastrophic {AI} risks}.
\newblock \bibinfo{journal}{\emph{arXiv preprint arXiv:2306.12001}}
  (\bibinfo{year}{2023}).
\newblock


\bibitem[Herbert(1965)]%
        {herbert1999dune}
\bibfield{author}{\bibinfo{person}{Frank Herbert}.}
  \bibinfo{year}{1965}\natexlab{}.
\newblock \bibinfo{booktitle}{\emph{Dune}}. Vol.~\bibinfo{volume}{1}.
\newblock \bibinfo{publisher}{Chilton Books}.
\newblock


\bibitem[Herbert(1976)]%
        {herbert2020children}
\bibfield{author}{\bibinfo{person}{Frank Herbert}.}
  \bibinfo{year}{1976}\natexlab{}.
\newblock \bibinfo{booktitle}{\emph{Children of dune}}.
  Vol.~\bibinfo{volume}{3}.
\newblock \bibinfo{publisher}{Putnam}.
\newblock


\bibitem[Jones and Bergen(2023)]%
        {jones2023does}
\bibfield{author}{\bibinfo{person}{Cameron Jones} {and}
  \bibinfo{person}{Benjamin Bergen}.} \bibinfo{year}{2023}\natexlab{}.
\newblock \showarticletitle{Does GPT-4 Pass the Turing Test?}
\newblock \bibinfo{journal}{\emph{arXiv preprint arXiv:2310.20216}}
  (\bibinfo{year}{2023}).
\newblock


\bibitem[Kahneman(2011)]%
        {kahneman2011thinking}
\bibfield{author}{\bibinfo{person}{Daniel Kahneman}.}
  \bibinfo{year}{2011}\natexlab{}.
\newblock \bibinfo{booktitle}{\emph{Thinking, fast and slow}}.
\newblock \bibinfo{publisher}{Macmillan}.
\newblock


\bibitem[Katz et~al\mbox{.}(2024)]%
        {katz2024gpt}
\bibfield{author}{\bibinfo{person}{Daniel~Martin Katz},
  \bibinfo{person}{Michael~James Bommarito}, \bibinfo{person}{Shang Gao}, {and}
  \bibinfo{person}{Pablo Arredondo}.} \bibinfo{year}{2024}\natexlab{}.
\newblock \showarticletitle{{GPT-4} passes the bar exam}.
\newblock \bibinfo{journal}{\emph{Philosophical Transactions of the Royal
  Society A}} \bibinfo{volume}{382}, \bibinfo{number}{2270}
  (\bibinfo{year}{2024}), \bibinfo{pages}{20230254}.
\newblock


\bibitem[Klofstad et~al\mbox{.}(2012)]%
        {klofstad2012sounds}
\bibfield{author}{\bibinfo{person}{Casey~A Klofstad}, \bibinfo{person}{Rindy~C
  Anderson}, {and} \bibinfo{person}{Susan Peters}.}
  \bibinfo{year}{2012}\natexlab{}.
\newblock \showarticletitle{Sounds like a winner: Voice pitch influences
  perception of leadership capacity in both men and women}.
\newblock \bibinfo{journal}{\emph{Proceedings of the Royal Society B:
  Biological Sciences}} \bibinfo{volume}{279}, \bibinfo{number}{1738}
  (\bibinfo{year}{2012}), \bibinfo{pages}{2698--2704}.
\newblock


\bibitem[Liu et~al\mbox{.}(2024)]%
        {liu2024conspemollm}
\bibfield{author}{\bibinfo{person}{Zhiwei Liu}, \bibinfo{person}{Boyang Liu},
  \bibinfo{person}{Paul Thompson}, \bibinfo{person}{Kailai Yang},
  \bibinfo{person}{Raghav Jain}, {and} \bibinfo{person}{Sophia Ananiadou}.}
  \bibinfo{year}{2024}\natexlab{}.
\newblock \showarticletitle{ConspEmoLLM: Conspiracy Theory Detection Using an
  Emotion-Based Large Language Model}.
\newblock \bibinfo{journal}{\emph{arXiv preprint arXiv:2403.06765}}
  (\bibinfo{year}{2024}).
\newblock


\bibitem[Nourooz~Pour(2023)]%
        {nourooz2023transitional}
\bibfield{author}{\bibinfo{person}{Hesam Nourooz~Pour}.}
  \bibinfo{year}{2023}\natexlab{}.
\newblock \showarticletitle{Transitional justice and online social platforms:
  Facebook and the Rohingya genocide}.
\newblock \bibinfo{journal}{\emph{International Journal of Law and Information
  Technology}} \bibinfo{volume}{31}, \bibinfo{number}{2}
  (\bibinfo{year}{2023}), \bibinfo{pages}{95--113}.
\newblock


\bibitem[Owens et~al\mbox{.}(2022)]%
        {owens2022exploring}
\bibfield{author}{\bibinfo{person}{Kentrell Owens}, \bibinfo{person}{Johanna
  Gunawan}, \bibinfo{person}{David Choffnes}, \bibinfo{person}{Pardis
  Emami-Naeini}, \bibinfo{person}{Tadayoshi Kohno}, {and}
  \bibinfo{person}{Franziska Roesner}.} \bibinfo{year}{2022}\natexlab{}.
\newblock \showarticletitle{Exploring deceptive design patterns in voice
  interfaces}. In \bibinfo{booktitle}{\emph{Proceedings of the 2022 European
  Symposium on Usable Security}}. \bibinfo{pages}{64--78}.
\newblock


\bibitem[Pinhanez et~al\mbox{.}(2024)]%
        {pinhanez2024creating}
\bibfield{author}{\bibinfo{person}{Claudio Pinhanez}, \bibinfo{person}{Raul
  Fernandez}, \bibinfo{person}{Marcelo Grave}, \bibinfo{person}{Julio Nogima},
  {and} \bibinfo{person}{Ron Hoory}.} \bibinfo{year}{2024}\natexlab{}.
\newblock \showarticletitle{Creating an African American-Sounding TTS:
  Guidelines, Technical Challenges, and Surprising Evaluations}.
\newblock \bibinfo{journal}{\emph{arXiv preprint arXiv:2403.11209}}
  (\bibinfo{year}{2024}).
\newblock


\bibitem[Pisanski et~al\mbox{.}(2018)]%
        {pisanski2018voice}
\bibfield{author}{\bibinfo{person}{Katarzyna Pisanski}, \bibinfo{person}{Anna
  Oleszkiewicz}, \bibinfo{person}{Justyna Plachetka}, \bibinfo{person}{Marzena
  Gmiterek}, {and} \bibinfo{person}{David Reby}.}
  \bibinfo{year}{2018}\natexlab{}.
\newblock \showarticletitle{Voice pitch modulation in human mate choice}.
\newblock \bibinfo{journal}{\emph{Proceedings of the Royal Society B}}
  \bibinfo{volume}{285}, \bibinfo{number}{1893} (\bibinfo{year}{2018}),
  \bibinfo{pages}{20181634}.
\newblock


\bibitem[Russell(2019)]%
        {russell2019human}
\bibfield{author}{\bibinfo{person}{Stuart Russell}.}
  \bibinfo{year}{2019}\natexlab{}.
\newblock \bibinfo{booktitle}{\emph{Human compatible: AI and the problem of
  control}}.
\newblock \bibinfo{publisher}{Penguin Uk}.
\newblock


\bibitem[Salvi et~al\mbox{.}(2024)]%
        {salvi2024conversational}
\bibfield{author}{\bibinfo{person}{Francesco Salvi},
  \bibinfo{person}{Manoel~Horta Ribeiro}, \bibinfo{person}{Riccardo Gallotti},
  {and} \bibinfo{person}{Robert West}.} \bibinfo{year}{2024}\natexlab{}.
\newblock \showarticletitle{On the conversational persuasiveness of large
  language models: A randomized controlled trial}.
\newblock \bibinfo{journal}{\emph{arXiv preprint arXiv:2403.14380}}
  (\bibinfo{year}{2024}).
\newblock


\bibitem[Shapiro(2023)]%
        {shapiro2023fancy}
\bibfield{author}{\bibinfo{person}{Scott Shapiro}.}
  \bibinfo{year}{2023}\natexlab{}.
\newblock \bibinfo{booktitle}{\emph{Fancy Bear Goes Phishing: The Dark History
  of the Information Age, in Five Extraordinary Hacks}}.
\newblock \bibinfo{publisher}{Random House}.
\newblock


\bibitem[Shedroff and Noessel(2012)]%
        {shedroff2012make}
\bibfield{author}{\bibinfo{person}{Nathan Shedroff} {and}
  \bibinfo{person}{Christopher Noessel}.} \bibinfo{year}{2012}\natexlab{}.
\newblock \bibinfo{booktitle}{\emph{Make it so: interaction design lessons from
  science fiction}}.
\newblock \bibinfo{publisher}{Rosenfeld Media}.
\newblock


\end{thebibliography}

\end{document}